\documentclass[useAMS,usenatbib]{mn2e}
\usepackage{graphics, graphicx}

\newif\ifAMStwofonts

\def\lapp{\ifmmode\stackrel{<}{_{\sim}}\else$\stackrel{<}{_{\sim}}$\fi}
\def\gapp{\ifmmode\stackrel{>}{_{\sim}}\else$\stackrel{>}{_{\sim}}$\fi}

\title[Multi-frequency profiles of pulsars]
{Multi-frequency integrated profiles of pulsars}
\author[Johnston et al.]
{Simon Johnston$^1$, Aris Karastergiou$^2$, Dipanjan Mitra$^3$
and Yashwant Gupta$^3$\\
$^1$Australia Telescope National Facility, CSIRO, P.O. Box 76, 
Epping, NSW 1710, Australia.\\
$^2$Astrophysics, University of Oxford, Denys Wilkinson Building, Keble
Road, OX1 3RH, Oxford, UK.\\
$^3$National Centre for Radio Astrophysics, TIFR, Pune University Campus,
Pune 411007, India.
}
\date{\today}
\pagerange{\pageref{firstpage}--\pageref{lastpage}}
\pubyear{2005}
\begin{document}
\maketitle
\label{firstpage}

\begin{abstract}
We have observed a total of 67 pulsars at five frequencies ranging
from 243 to 3100~MHz. Observations at the lower frequencies were made
at the Giant Metre Wave Telescope in India and those at higher frequencies
at the Parkes Telescope in Australia. We present profiles from 34 of
the sample with the best signal to noise ratio and the least scattering.
The general `rules' of pulsar profiles are seen in the data; profiles
get narrower, the polarization fraction declines and outer components 
become more prominent as the frequency increases. Many counterexamples
to these rules are also observed, and pulsars with complex profiles
are especially prone to rule breaking. We hypothesise that the location
of pulsar emission within the magnetosphere evolves with time as the
the pulsar spins down. In highly energetic pulsars, the emission
comes from a confined range of high altitudes, in the middle range
of spin down energies the emission occurs over a wide range of 
altitudes whereas in pulsars with low spin-down energies it is confined 
to low down in the magnetosphere.
\end{abstract}

\begin{keywords}
pulsars:general
\end{keywords}

\section{Introduction}
The frequency dependence of radio emission from pulsars is a key
aspect of the emission mechanism.
In the radio part of the electromagnetic spectrum, pulsars are known
to emit from as low as 10~MHz (Erickson \& Mahoney 1985\nocite{em85})
to at least 150~GHz (Camilo et al. 2007\nocite{crp+07}).
Radio emission in pulsars is thought to originate from the
ultra-relativistic plasma which is streaming out along a group of
open magnetic field lines centered around the magnetic poles. The frequency of
emission is naturally related to the local plasma conditions, and the
frequency dependence of the properties of pulsar emission can
be used to probe the active magnetosphere. Since it is not
unreasonable to expect that the local plasma conditions change with
height above the pulsar surface, it is to some degree expected that
observations at different frequencies sample different parts
of the open field line region.

It is possible to probe the magnetosphere in two different ways.
First, by looking at individual pulses over a wide range of
frequencies one can examine the instantaneous, fluctuating properties
of the plasma emission (e.g. Karastergiou et al. 2002\nocite{kkj+02},
Bhat et al. 2007\nocite{bgk+07}). Secondly, one can determine
the long term structure of the magnetic field, the average properties
of the magnetic field and the star's geometry through a study
of the time-averaged profiles. 
These time averaged
profiles of pulsars show frequency dependent characteristics which,
over the years, have aided our understanding in determining where
in the magnetosphere the radio emission arises.
Some of these
characteristics are: (i) pulsar emission has a steep spectral index with a mean
spectral index of $\sim-1.6$ and a large spread in values from 0 to
$\sim-4$ (Maron et al. 2000\nocite{mkk+00}),
(ii) pulse profiles appear to widen as the frequency decreases
with the implication that the lower frequencies are emitted from higher
in the magnetosphere where the flaring of the magnetic field lines
is more pronounced (Thorsett 1991, Mitra \& Rankin 2002\nocite{tho91a,mr02a}),
(iii) different components of the pulse profile
have different spectral indices with the more central components appearing
to have a steeper spectrum that the outer components. As a consequence,
low frequency profiles are often dominated by a single component whereas
these components can be flanked by outriders at higher frequencies
(Backer 1976, Rankin 1983\nocite{bac76,ran83}),
(iv) the overall polarization level generally decreases with increasing
frequency although in some components polarization can increase.

The most comprehensive multi-frequency observations of pulse profiles
were carried out by \cite{gl98} at frequencies between 234
and 1650~MHz. This database can be supplemented by observations at
100~MHz by the Russian groups (e.g. Izvekova, Malofeev \& Shitov 1989;
Kuzmin \& Losovskii 1999\nocite{ims89,kl99b})
and at higher frequencies by the Effelsberg (e.g. Kramer et al. 1997;
von Hoensbroech \& Xilouris 1997)\nocite{kxj+97,hx97} and
Parkes (Karastergiou, Johnston \& Manchester 2005;
Johnston, Karastergiou \& Willett 2006)\nocite{kjm05,jkw06} groups.

These observations have been put to good use in constructing beam
models for radio pulsars, most notably by Rankin and colleagues
(Rankin 1983; Rankin 1993; Mitra \& Deshpande 1999; Mitra \& Rankin 2002
\nocite{ran83,ran93,md99,mr02a}) and Lyne \& Manchester (1988)\nocite{lm88}.
The former authors postulate
a beam which consists of one or more concentric rings of emission
located about the magnetic pole which also produces emission.
The latter authors, in contrast, prefer a model where emission regions
are located randomly across the polar cap.
More recently, \cite{kj07} have produced an alternative model in which
the emission is produced at multiple heights (at a given frequency)
and is confined to a region near the last open field lines.

In this paper we present polarization profiles at 5 frequencies (243, 322, 690,
1400 and 3100~MHz) for 34 pulsars, most of which have not been
previously observed at the upper and lower frequency end.
Section 2 outlines our source selection.
The observations were made with the Giant Metre Wave Telescope (GMRT)
in India and the Parkes radio telescope in Australia and these are
described in detail in Section 3. In Section 4 we present the results
and discuss their implications in terms of current models in Section 5.

\section{Source Selection}
We have embarked on a campaign to produce a database of integrated
pulse profiles in full polarization and with high time resolution.
At present we have observed close to 400 pulsars at frequencies above
0.6~GHz using the Parkes radio telescope. Profiles from a selection
of these pulsars have already been published (Johnston et al. 2005;
Karastergiou \& Johnston 2006; Johnston et al. 2006;
Johnston et al. 2007; Weltevrede \& Johnston 2008
\nocite{jhv+05,kj06,jkw06,jkk+07,wj08}).
As the Parkes telescope
does not have a good receiver at frequencies below 0.6~GHz, we supplemented
the Parkes observations with low frequency observations at the
Giant Metre Wave Telescope (GMRT). We selected sources to observe at
low frequencies with the following criteria. First, the pulsars should
have declinations northwards of $-50$\degr\ in order to be visible at
the GMRT and southwards of +27\degr\ to be visible at Parkes.
Secondly, the pulsars must have low dispersion and scattering
measures in avoid smearing of the profiles at low frequencies. Finally
we needed to choose high flux density pulsars far from the Galactic plane
to overcome the high sky brightness temperatures at 200~MHz and the
low fluxes of pulsars at high frequencies.
This resulted in a sample of 67 pulsars.

\section{Observations and Data Analysis}
\subsection{GMRT Observations}
The GMRT observations were carried out in 
2007 September spread over 3 days for a total of 30~h.
This yielded single pulse polarimetric data for 67 pulsars observed 
simultaneously at 243, 322 and 607 MHz. 

The GMRT is an interferometric array consisting 
of 30 antennas, each of 45 m diameter,
and operates at six different wavebands between 150 and
1450 MHz (Swarup et al. 1991\nocite{sak+91}).
The array is Y shaped, populated with  
14 central antennas in a region of about 1~km$^{2}$, and the
remaining 16 are spread along three arms with a longest baseline of 25 km. 
Polarimetric pulsar observations
at the GMRT are done using the phased array mode of operation
(see Sirothia 2000, Gupta et al. 2000\nocite{sir00,ggj+00})
where the signals from various antennas are added coherently or in phase.
Subsets of 30 antennas can 
be grouped together to form subarrays operating at different wavebands
which gives the flexibility of observing simultaneously at multiple 
frequencies. The details of the simultaneous, multifrequency, phased array
observing technique 
can be found in Ahuja et al. (2005), Smits et al (2007) and 
in Bhattacharyya et al. (2007)\nocite{agmk05,sms+07,bgg08}. Here we briefly 
describe the specifics of the scheme we have employed for our observations.

For our purpose we used three subarrays, with each subarray operating 
at a centre frequency of 243, 322 and 607 MHz respectively.
To minimise the effect of spatial variation of 
the ionosphere the antennas chosen for 243 MHz were comprised of 8
closely spaced central antennas, for 322 MHz 6 central antennas
and one arm antenna were used and for 610 MHz 9 arm antennas were chosen.
Each of these wavebands has dual linear feeds
which are converted to left and right hand circular via a hybrid.
Subsequently two polarization channels 
of 32 MHz bandwidth are downconverted
and further split into upper and lower sideband (USB and LSB) of 16 MHz.
The orthogonally polarized complex voltages
are then sampled at the Nyquist rate at each antenna.
The digitally sampled signals of 16 MHz per polarization channel per sideband 
is divided into 256 channels by an FX correlator and 
are added in phase for each polarization.
These outputs are fed to the pulsar backend which computes 
the auto-and cross-polarized power, and is resampled at a rate of 0.512 ms.
The backend can be configured to set different values of gains to 
each spectral channel of each antenna and each polarization before the 
signals are added using a technique known as bandmasking
(see Bhattacharyya et al. 2007\nocite{bgg08} for details).
We have this flexibility to assign
the first 110 channels (6.87 MHz) of the USB to the 243 MHz subarray, 
channels 115 - 256 (8.8 MHz) of the USB to the 322 MHz subarray,
and the whole 256 channels of the LSB to the 607 MHz subarray.
The final time series output has properties equivalent to data recorded
using a large single dish and similar
polarization calibration methods can be applied 
to get calibrated Stokes parameters
(Johnston 2002, Mitra et al. 2005\nocite{joh02}).

During the observations we did an initial phasing with respect to a 
reference antenna of each subarray on a strong calibrator source.
For observing pulsars, a secondary calibrator source close to the pulsar 
was used to validate the phasing and if needed 
phasing was redone. Phasing was done at 1$-$2 hour
intervals.

The recorded data were analysed 
using polarization analysis pulsar software developed at the GMRT. 
The raw auto-and cross-polarized power were first gain corrected and 
used to obtain the measured Stokes parameters for the individual spectral 
channels. For collapsing the channels the fixed delay of the two 
polarization channels of the reference antenna was corrected 
along with the pulsar's dispersion and rotation measure.
Observations were made each day of the bright pulsar PSR~B1929+10 at
a variety of different parallactic angles. From these data, we estimate the
error on the linear and circular polarization to be $\sim$5 per cent.

\subsection{Parkes Observations}
\begin{table*}
\caption{Information on the 34 pulsars observed with good signal to noise
ratio and low scattering at 243, 327, 690, 1400 and 3100 MHz.}
\begin{tabular}{llcccrrrrr}
\hline & \vspace{-3mm} \\
PSR J & PSR B & \multicolumn{1}{c}{Period} & \multicolumn{1}{c}{log($\dot{E}$)}
& \multicolumn{1}{c}{DM} & \multicolumn{5}{c}{10\% Width (degrees)} \\
& & \multicolumn{1}{c}{(s)} & \multicolumn{1}{c}{(erg~s$^{-1}$)}
& \multicolumn{1}{c}{(cm$^{-3}$pc)} 
& 243 & 327 & 690 & 1400 & 3100 \\
\hline & \vspace{-3mm} \\
J0034$-$0721 &  B0031$-$07 & 0.9429 & 31.3 &  11.38 & 41.8 & 41.1 & 41.8 & 36.6 & 32.0\\
J0134$-$2937 &             & 0.1369 & 33.1 &  21.81 & 35.1 & 39.1 & 25.0 & 17.9 & 22.9\\
J0151$-$0635 &  B0148$-$06 & 1.4646 & 30.7 &  25.66 & 50.6 & 41.5 & 45.0 & 42.9 & 40.1\\
J0152$-$1637 &  B0149$-$16 & 0.8327 & 31.9 &  11.92 & 11.7 & 11.1 & 11.3 &  9.8 & 10.2\\
J0304+1932   &  B0301+19   & 1.3875 & 31.3 &  15.74 & 21.6 & 21.4 & 17.0 & 15.1 & 14.4\\
J0525+1115   &  B0523+11   & 0.3544 & 31.8 &  79.34 & 27.5 & 23.9 & 17.6 & 17.2 & 18.9\\
J0543+2329   &  B0540+23   & 0.2459 & 34.6 &  77.71 & 30.0 & 30.0 & 23.6 & 24.3 & 23.6\\
J0614+2229   &  B0611+22   & 0.3349 & 34.8 &  96.91 & 21.1 & 17.6 & 13.4 & 14.4 & 13.0\\
J0630$-$2834 &  B0628$-$28 & 1.2444 & 32.2 &  34.47 & 38.1 & 38.0 & 38.7 & 34.4 & 30.6\\
J0729$-$1836 &  B0727$-$18 & 0.5101 & 33.7 &  61.29 & 24.2 & 22.4 & 20.4 & 19.4 & 18.9\\
J0837+0610   &  B0834+06   & 1.2737 & 32.1 &  12.89 &  8.4 &  8.6 &  9.1 &  9.1 &  9.5\\
J0908$-$1739 &  B0906$-$17 & 0.4016 & 32.6 &  15.89 & 23.0 & 20.2 & 20.0 & 20.7 & 21.2\\
J0922+0638   &  B0919+06   & 0.4306 & 33.8 &  27.27 & 22.7 & 19.2 & 16.9 & 14.8 &  8.4\\
J1507$-$4352 &  B1504$-$43 & 0.2867 & 33.4 &  48.7  & 11.9 & 12.7 & 14.0 & 13.4 & 10.2\\
J1559$-$4438 &  B1556$-$44 & 0.2570 & 33.4 &  56.1  & 50.2 & 22.2 & 16.5 & 25.3 & 28.1\\
J1645$-$0317 &  B1642$-$03 & 0.3876 & 33.1 &  35.73 &  6.2 &  6.2 &  6.3  & 7.4 & 15.1\\
J1703$-$3241 &  B1700$-$32 & 1.2117 & 31.2 & 110.31 & 24.6 & 17.1 & 14.8 & 14.1 & 13.4\\
J1705$-$1906 &  B1702$-$19 & 0.2989 & 33.8 &  22.91 & 19.7 & 17.6 & 17.2 & 16.9 & 16.6\\
J1709$-$1640 &  B1706$-$16 & 0.6530 & 32.9 &  24.87 & 12.7 & 12.3 & 13.0 & 12.0 &  8.1\\
J1731$-$4744 &  B1727$-$47 & 0.8298 & 34.0 & 123.33 & 15.1 & 12.2 & 10.5 & 10.2 & 10.9\\
J1733$-$2228 &  B1730$-$22 & 0.8716 & 30.4 &  41.14 & 17.3 & 34.8 & 41.1 & 37.3 & 40.8\\
J1735$-$0724 &  B1732$-$07 & 0.4193 & 32.8 &  73.51 & 13.4 & 10.5 & 16.5 & 20.4 & 19.7\\
J1740+1311   &  B1737+13   & 0.8030 & 32.0 &  48.67 & 20.7 & 20.7 & 21.8 & 25.3 & 24.3\\
J1745$-$3040 &  B1742$-$30 & 0.3674 & 33.9 &  88.37 & 36.6 & 26.7 & 19.3 & 21.2 & 22.9\\
J1825$-$0935 &  B1822$-$09 & 0.7689 & 33.7 &  19.38 &  9.5 &  9.8 & 21.1 & 21.1 & 21.4\\
J1844+1454   &  B1842+14   & 0.3754 & 33.1 &  41.51 & 12.0 & 12.7 & 24.3 & 25.0 & 16.9\\
J1850+1335   &  B1848+13   & 0.3455 & 33.1 &  60.15 & 14.8 & 16.9 & 12.0 & 12.0 & 11.3\\
J1900$-$2600 &  B1857$-$26 & 0.6122 & 31.5 &  37.99 & 26.7 & 39.4 & 38.7 & 41.1 & 35.2\\
J1913$-$0440 &  B1911$-$04 & 0.8259 & 32.5 &  89.38 &  7.4 &  6.0 &  6.7 &  8.1 &  8.4\\
J1921+2153   &  B1919+21   & 1.3373 & 31.3 &  12.46 & 10.5 & 10.9 & 11.3 & 12.7 & 11.6\\
J1941$-$2602 &  B1937$-$26 & 0.4028 & 32.8 &  50.04 & 10.5 & 12.4 &  9.8 & 12.0 &  9.8\\
J2048$-$1616 &  B2045$-$16 & 1.9615 & 31.8 &  11.46 & 18.4 & 17.6 & 16.2 & 14.8 & 13.7\\
J2116+1414   &  B2113+14   & 0.4401 & 32.1 &  56.15 & 16.9 & 16.2 & 16.2 & 18.3 & 19.0\\
J2330$-$2005 &  B2327$-$20 & 1.6436 & 31.6 &   8.46 &  8.0 &  7.4 &  7.0 &  6.7 &  6.3\\
\hline & \vspace{-3mm} \\
\end{tabular}
\label{sources}
\end{table*}

We carried out the observations with the Parkes radio telescope
in 2006 August in two separate observing sessions followed by a 
supplementary session in 2007 December.
This resulted in polarization data on more than 250 pulsars
with high signal to noise ratio at frequencies of 0.67, 1.4
and 3.1~GHz.

In the first session which took place from 2006 Aug 14 to 17 we used a
dual frequency receiver system capable of observing simultaneously in both
the 50 and 10~cm bands.  We used central frequencies of 3.1~GHz with
a bandwidth of 512~MHz and a resolution of 0.5~MHz and 0.69~GHz with an 
effective bandwidth (after interference rejection) of 35~MHz and
a resolution of 0.125~MHz.  On-line folding with a digital
correlator system resulted in a total of 1024 phase bins
per pulse period for each Stokes parameter. Data were recorded to disk
at 30~s intervals for a typical observing period of 30~m.

The second session from 2006 August 24 to 27 was conducted with
a centre frequency of 1368~MHz, using the H-OH receiver at the
prime focus of the Parkes telescope.  The total bandwidth was 512 MHz
subdivided into 1024 frequency channels in order to remove the
effects of interstellar dispersion.  Again, 1024 channels
were obtained across the band and 1024 phase bins per pulse period
per Stokes parameter recorded.

All receivers have orthogonal linear feeds and also have a pulsed
calibration signal which can be injected at a position angle of
45\degr\ to the two feed probes. An observation of the calibration signal
was made prior to every pulsar observation in order to calibrate
the polarization and the flux density.

Data analysis was carried out using the PSRCHIVE software package
(Hotan, van Straten \& Manchester 2004\nocite{hvm04})
and the analyis and calibration were carried out in an identical fashion
to that described in detail in \cite{jhv+05}.
Most importantly, we are able to determine absolute position angles for 
the linearly polarized radiation at all three of our observing frequencies.
We estimate the errors on the linear and circular polarization to
be $\sim$2 per cent.

\section{Results}
Table~\ref{sources} gives information on the 34 pulsars which had
good signal to noise ratio (s/n) at all frequencies and 
minimal scattering (less than 2\degr) at the 243~MHz.
As scattering is a strong function of observing frequency (with
a power-law index of $-4$), this implies that the scattering is
less than 1\degr, and therefore not significant, at the higher frequencies.
The table lists the pulsar name, period, spin-down energy ($\dot{E}$)
and dispersion measure followed
by the width of the profile in degrees measured at 10\% of 
the peak amplitude for each of the five frequencies.

Table~\ref{nondet} lists the 33 pulsars for which weak or no detections
were made or the profile was highly scattered so that component features
were blurred out at 322 and/or 243~MHz.
\begin{table}
\caption{Pulsars observed with the GMRT but not used in this sample.
Y and N denote detections and non-detections.}
\begin{tabular}{llccl}
\hline & \vspace{-3mm} \\
PSR J & PSR B & 243~MHz & 327~MHz & Comment \\
\hline & \vspace{-3mm} \\
J0108$-$1439 &            & N & Y \\
J0520$-$2553 &            & Y & Y & Low s/n \\
J0601$-$0572 & B0559$-$05 & Y & Y & Scattered \\
J0624$-$0424 & B0621$-$04 & Y & Y & Low s/n \\
J0631+1036   &            & N & N \\
J0659+1414   & B0656+14   & N & N \\
J0738$-$4042 & B0736$-$40 & N & Y & Scattered \\
J0742$-$2822 & B0740$-$28 & Y & Y & Scattered \\
J1514$-$4834 & B1510$-$48 & N & Y \\
J1535$-$4114 &            & N & N \\
J1536$-$3602 &            & N & N \\
J1549$-$4848 &            & N & Y \\
J1557$-$4258 &            & N & Y & Scattered \\
J1605$-$5257 & B1601$-$52 & N & N \\
J1641$-$2347 &            & N & N \\
J1700$-$3312 &            & N & N \\
J1740$-$3015 & B1737$-$30 & N & N \\
J1743$-$4212 &            & N & N \\
J1801$-$2920 & B1758$-$29 & N & Y \\
J1807$-$0847 & B1804$-$08 & N & Y & Scattered \\
J1808$-$0813 &            & N & N \\
J1820$-$0427 & B1818$-$04 & Y & Y & Scattered \\
J1822$-$2256 & B1819$-$22 & N & Y \\
J1835$-$1106 &            & N & N \\
J1848$-$0123 & B1845$-$01 & N & Y & Scattered \\
J1852$-$2610 &            & N & N \\
J1901$-$0906 &            & N & N \\
J1917+1353   & B1915+13   & Y & Y & Scattered \\
J1932+1059   & B1929+10   & Y & Y & Calibrator \\
J1935+1616   & B1933+16   & Y & Y & Scattered \\
J1937+2544   & B1935+25   & Y & Y & Low s/n \\
J2006$-$0807 & B2003$-$08 & N & Y \\
J2108$-$3429 &            & Y & Y & Low s/n \\
\hline & \vspace{-3mm} \\
\end{tabular}
\label{nondet}
\end{table}

\noindent
{\bf PSR~J0034$-$0721:} The integrated profile of this pulsar is
broader at lower frequencies than at higher. It appears that 
the trailing component gradually declines in intensity with increasing
frequency. The fractional polarization remains low at all frequencies.
The PA swing varies significantly with frequency. In particular
the location of the orthogonal mode jump shifts pulse phase as a 
function of frequency and is perhaps entirely absent at 4.8~GHz
(von Hoensbroech et al. 1998)\nocite{hkk98}. This is consistent with the
idea that the orthogonal modes have different spectral indices 
(Karastergiou et al. 2005)\nocite{kjm05}.
The geometry of this star has been discussed in detail by \cite{sms+07}.

\noindent
{\bf PSR~J0134$-$2937:} The frequency evolution of this pulsar is
interesting. At frequencies above 600~MHz, the trailing edge of the
profile is very sharply defined and the trailing component dominates
almost completely. In the 435~MHz profile of Manchester, Han and Qiao
(1998)\nocite{mhq98} the trailing edge is less
sharp and in our 325~MHz profile, not only has the trailing edge been
smoothed out, but an unpolarized leading component also appears.
The 243~MHz profile has low s/n but again the leading component is apparent.
The polarization is high below $\sim$1~GHz but declines thereafter.
As noted by Manchester et al. (1998), the PA swing is very shallow.
This appears to be a partial cone with the steepest swing of PA occur
ahead of the main component.

\noindent
{\bf PSR~J0151$-$0635:} The profile of this pulsar is a classic double
profile, with the two components joined through a saddle or bridge.
The profile narrows with increasing frequency and the leading component
has the flatter spectral index.  It appears as if the steep swing of 
PA occurs between the two components. At 243~MHz however, the PA traverse
is more complicated; the same is seen in the \cite{gl98} profile.
The linear polarization appears to first increase before then
decreasing in the leading component. As is common in many of these
double pulsars, the polarization fraction is low at the profile edges.
The width of the profile decreases with increasing frequency.

\noindent
{\bf PSR~J0152$-$1637:} This pulsar also has a double profile with
the components blended together. There is little evolution with 
frequency either in the profile width or the relative amplitudes of
the two components. The PA swing is complicated by two orthogonal jumps
near the leading part of the profile.

\noindent
{\bf PSR~J0304+1932:} The profile of this pulsar is very similar to
that of PSR~J0151$-$0635 except that the components are closer together.
There is a hint of a central component at the lowest frequencies.
There is a significant narrowing of the profile with increasing
frequency and the trailing component has the flatter spectral index.
Again the steepest swing of PA occurs in the centre of the profile
and indeed the PA swing is far less complex than that in either
PSR~J0151$-$0635 or PSR~J0152$-$1637. The pulse width decreases as
the frequency increases as does the fractional polarization.
The line of sight likely goes close to the magnetic axis 
(Mitra \& Li 2004\nocite{ml04}).

\noindent
{\bf PSR~J0525+1115:} At face value this pulsar appears to have two
prominent and symmetrical components at all frequencies with the
leading component having the steeper spectral index. There is a hint
of a more central component particularly at 1.4~GHz (see also
Weisberg et al. 1999\nocite{wcl+99}). Also, at the two lowest frequencies
there is perhaps a weak trailing component.
The circular polarization has a sign change under this central component
at 1.4~GHz and higher frequencies. Both \cite{wcl+99} and \cite{jhv+05}
took the location of the sign change to be the magnetic pole crossing.
However, at 660~MHz and 435~MHz (Weisberg et al. 2004)\nocite{wck+04}
the circular polarization is confined to the middle of the profile and has 
no sign change.  The linear polarization is at its maximum, with the 
steepest swing of PA occurring some 3\degr\ after the profile midpoint.
At still lower frequencies the linear polarization decreases again.
There is little, if any, decrease in the overall profile width over this
frequency range.
\begin{figure*}
\includegraphics[height=0.7\textheight,angle=0]{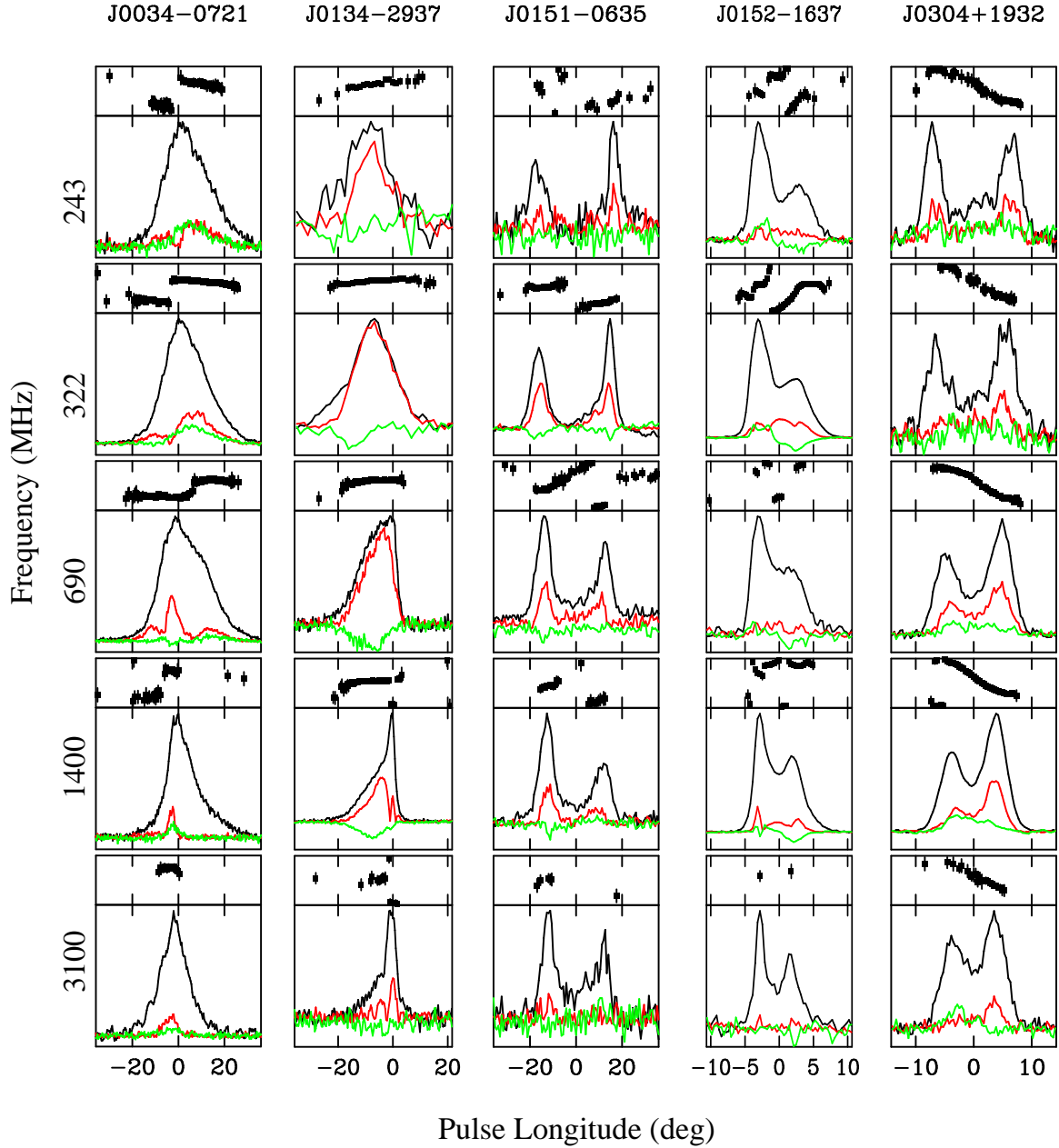}
\caption{Integrated profiles for 5 pulsars at 5 frequencies.
The pulsar name is indicated along the top. The plots show
the integrated pulse profiles as a function of longitude for
each pulsar. From top to bottom the frequencies are 243, 320, 660,
1400, 3100 MHz. Each plot is subdivided into two panels. The lower
plot shows the total intensity (black line), linear polarization
(red line) and circular polarization (green line) as a function
of pulse longitude. The top panel 
shows the position angle of the linearly polarized radiation and
runs between $-90$\degr\ and +90\degr.}
\end{figure*}
\begin{figure*}
\includegraphics[height=0.7\textheight,angle=0]{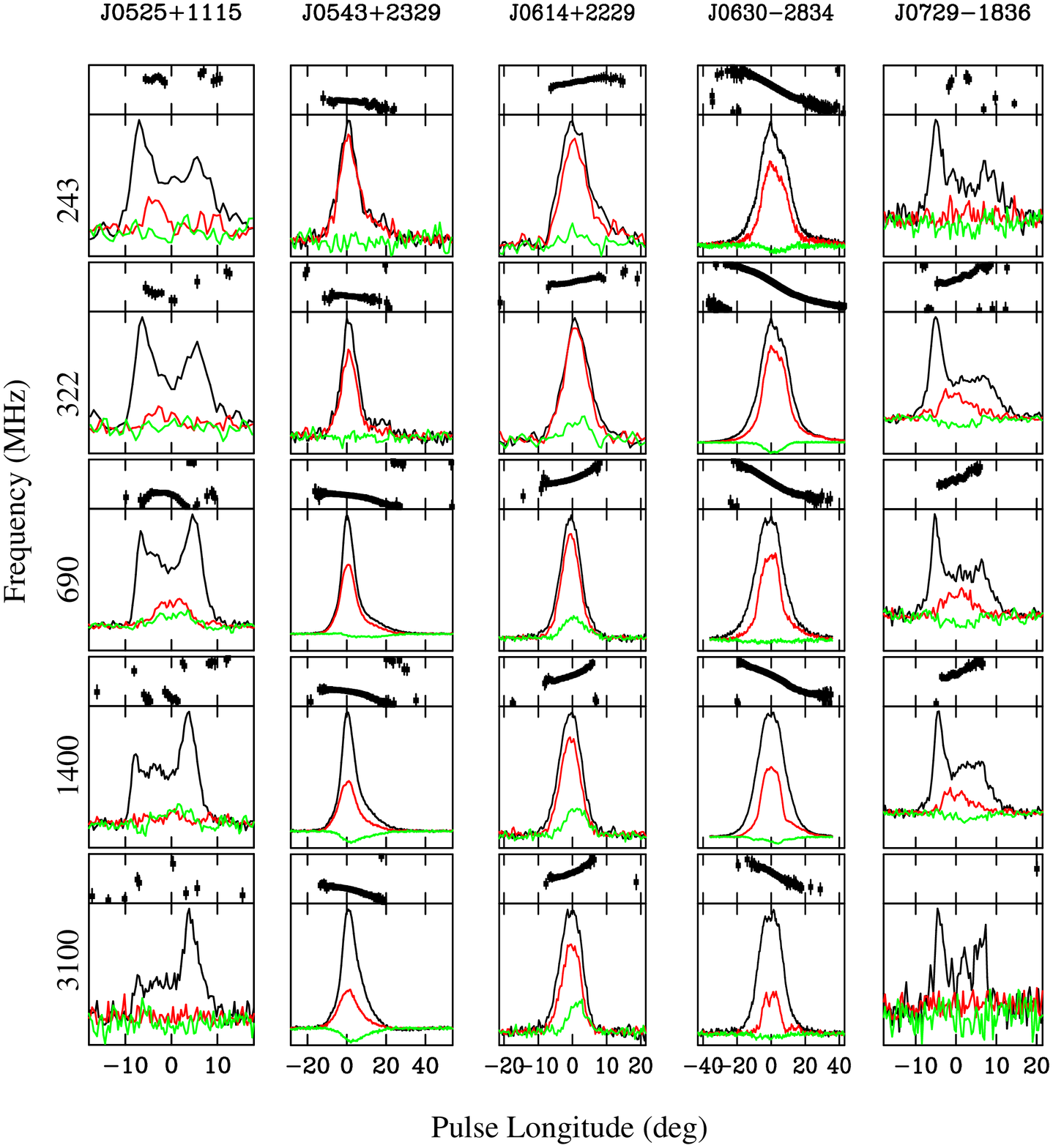}
\caption{Integrated profiles for 5 pulsars at 5 frequencies as marked.
See Figure~1 for details.}
\end{figure*}
\begin{figure*}
\includegraphics[height=0.7\textheight,angle=0]{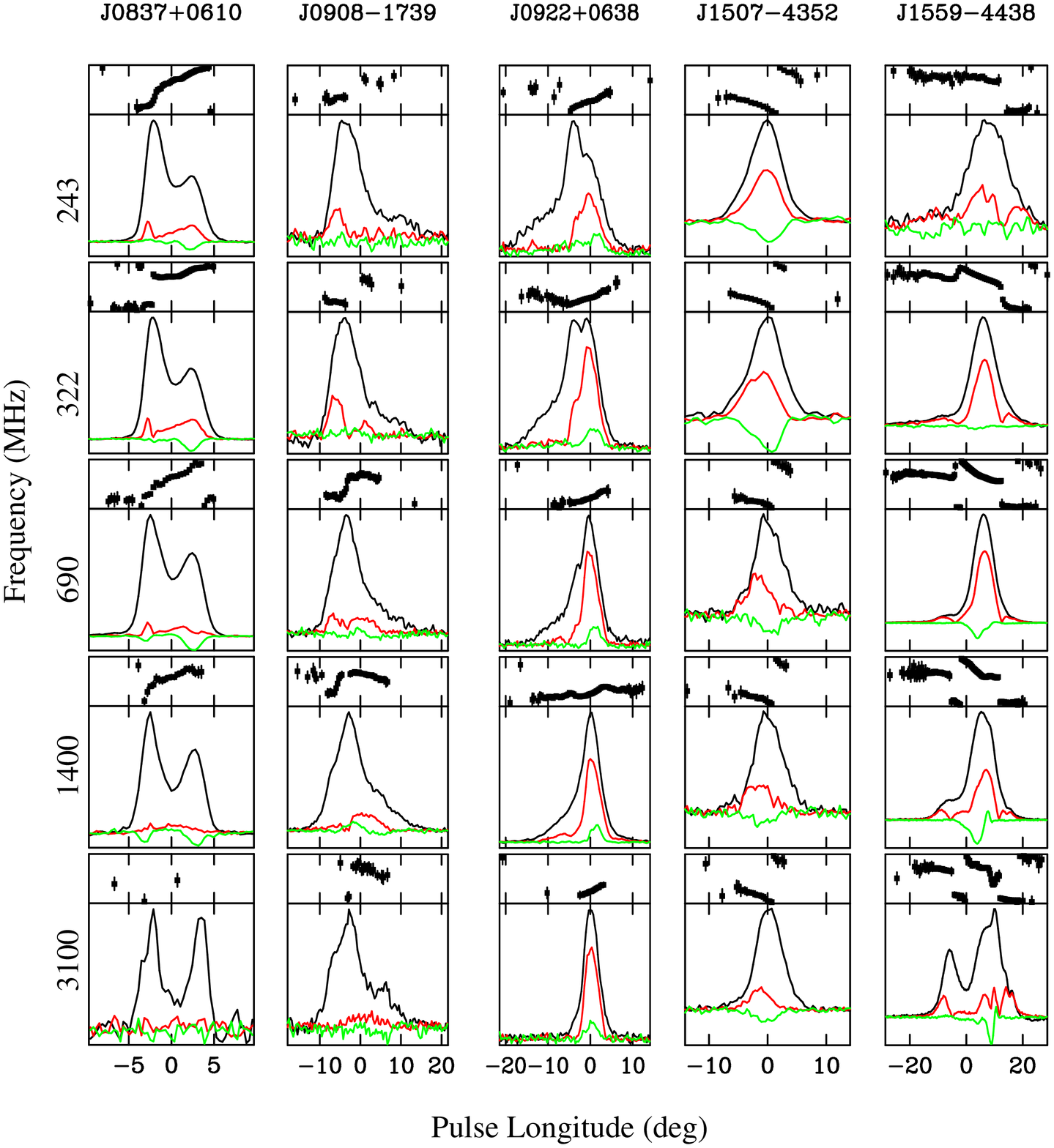}
\caption{Integrated profiles for 5 pulsars at 5 frequencies as marked.
See Figure~1 for details.}
\end{figure*}
\begin{figure*}
\includegraphics[height=0.7\textheight,angle=0]{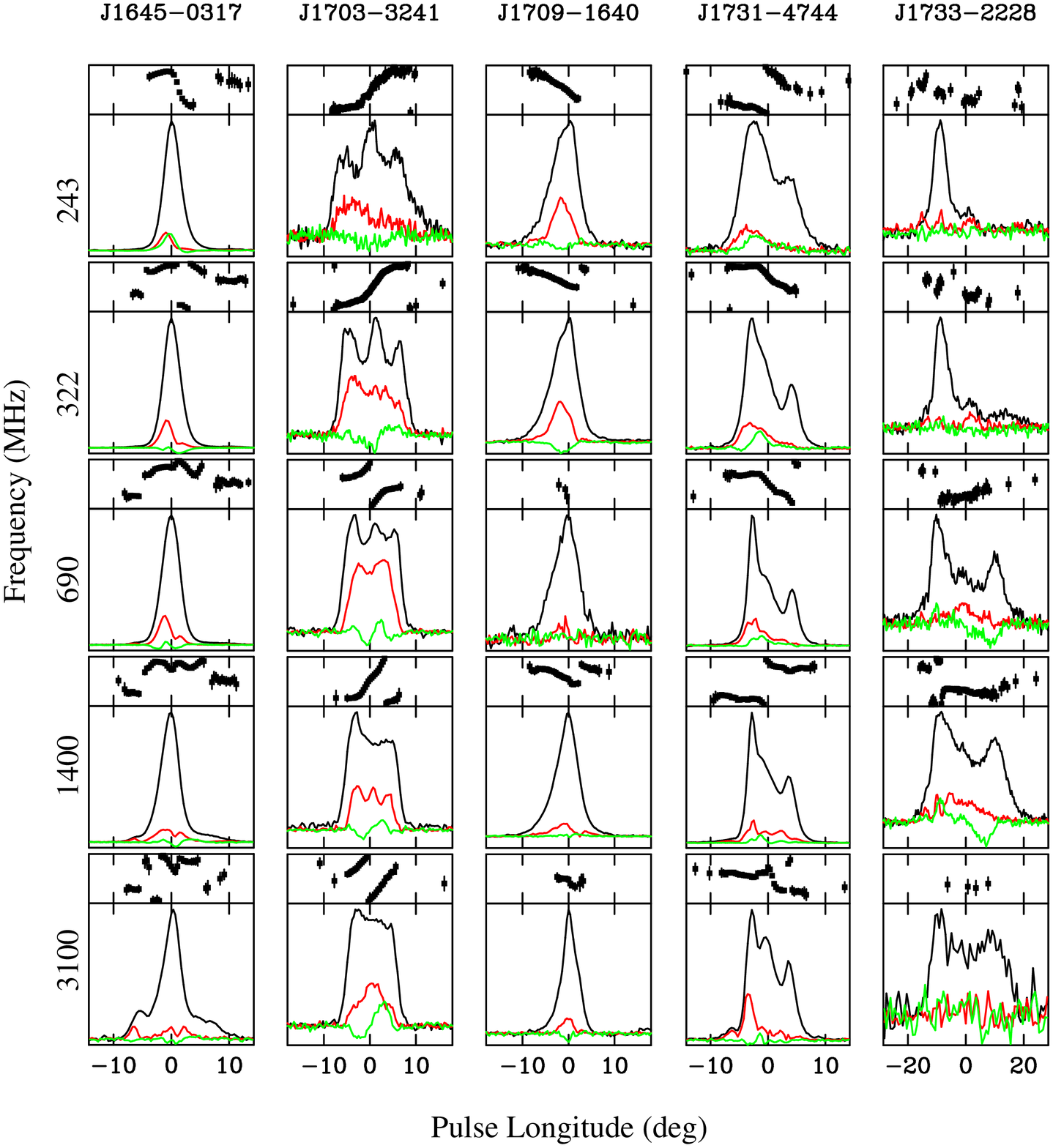}
\caption{Integrated profiles for 5 pulsars at 5 frequencies as marked.
See Figure~1 for details.}
\end{figure*}
\begin{figure*}
\includegraphics[height=0.7\textheight,angle=0]{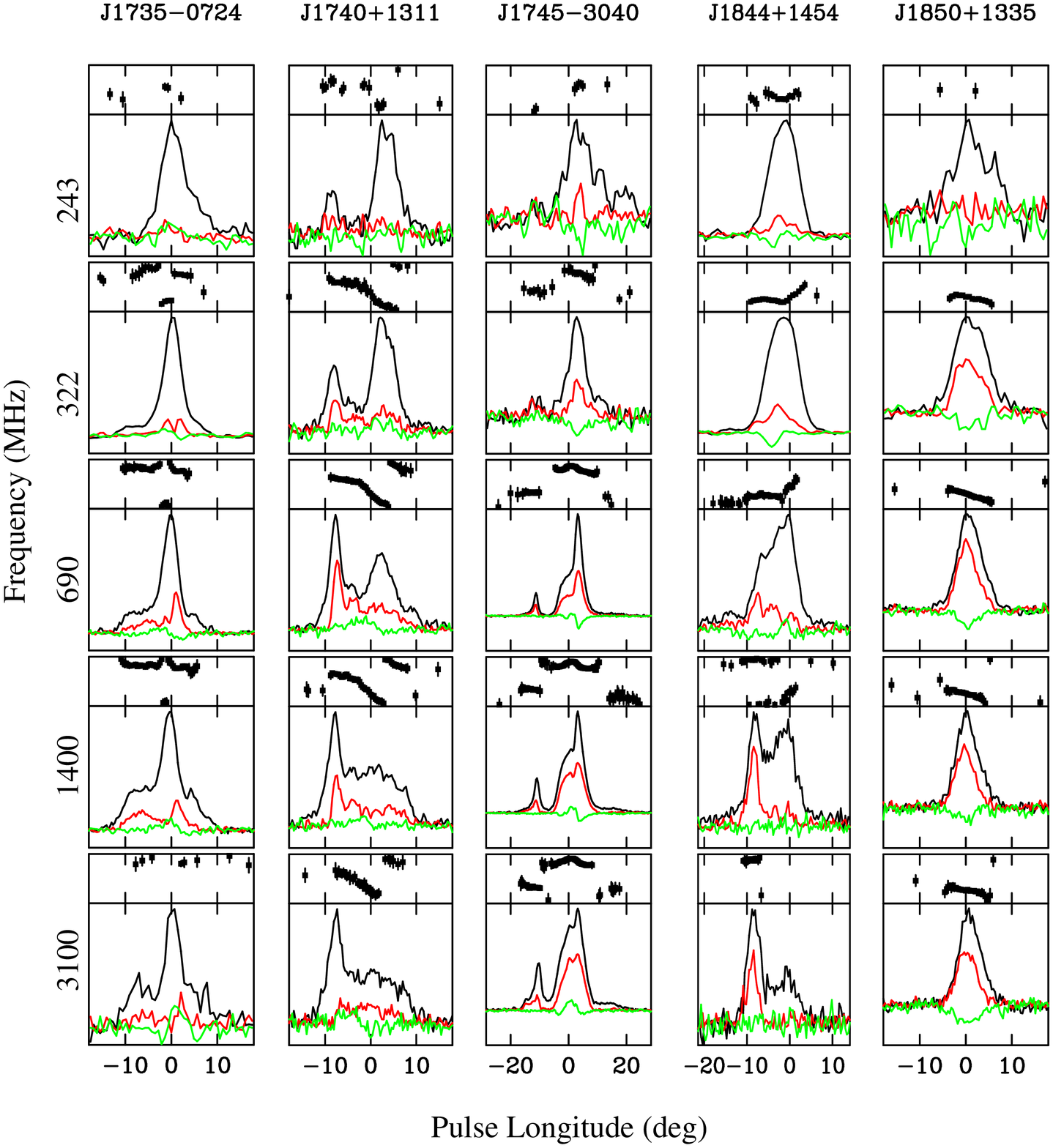}
\caption{Integrated profiles for 5 pulsars at 5 frequencies as marked.
See Figure~1 for details.}
\end{figure*}
\begin{figure*}
\includegraphics[height=0.7\textheight,angle=0]{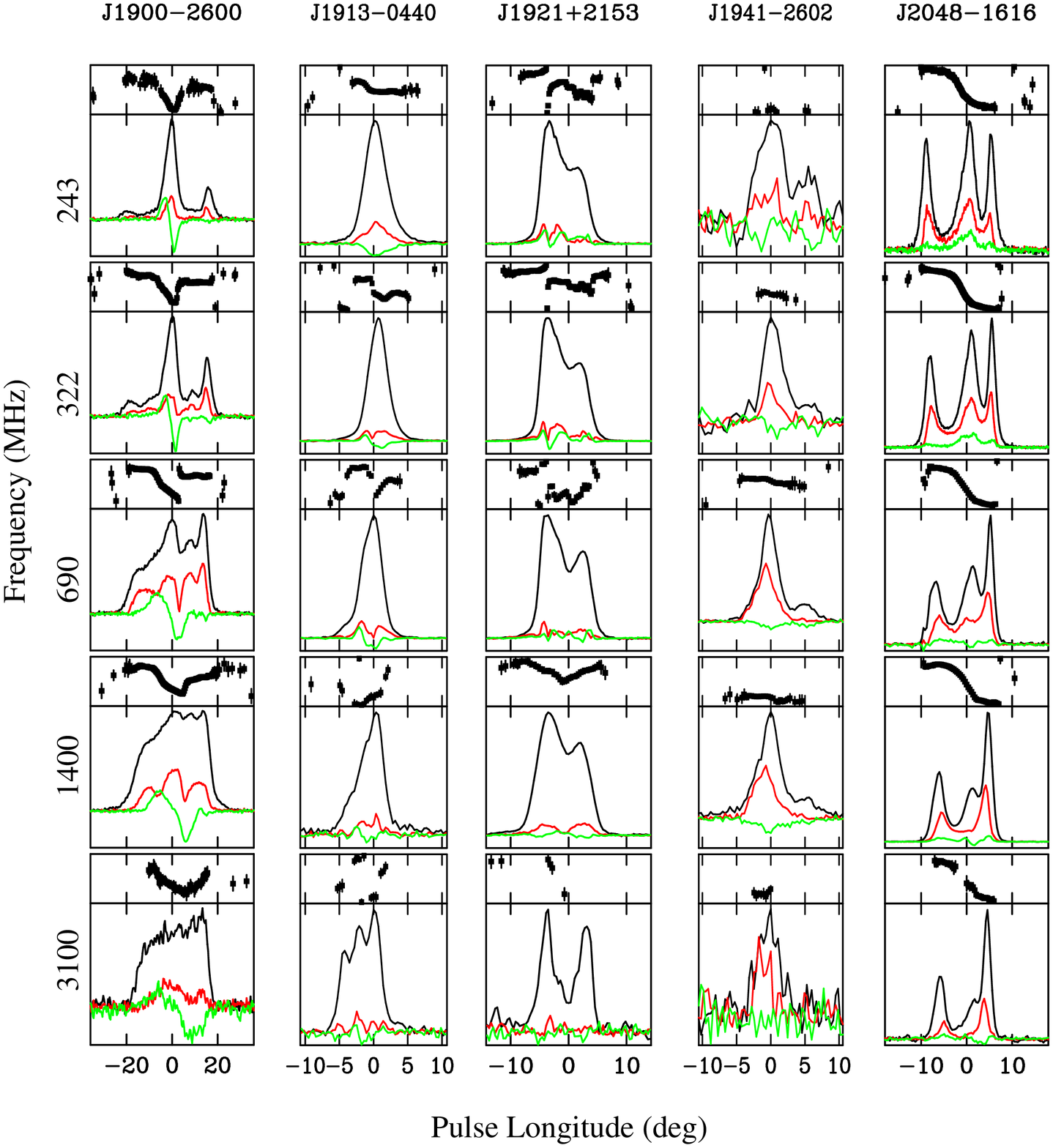}
\caption{Integrated profiles for 5 pulsars at 5 frequencies as marked.
See Figure~1 for details.}
\end{figure*}

\noindent
{\bf PSR~J0543+2329:} The profile from this pulsar appears to come from
the leading edge of the beam (Lyne \& Manchester 1988\nocite{lm88}).
The PA clearly steepens towards later
longitudes (see the discussion in Johnston et al. 2007\nocite{jkk+07}).
The profile width declines with frequency as the trailing edge
becomes sharper.
The main change is that the linear polarization decreases with increasing
frequency whereas the circular polarization slowly increases (as is seen in a 
number of other pulsars; see von Hoensbroech \& Lesch 1999\nocite{hl99}).

\noindent
{\bf PSR~J0614+2229:} This pulsar is similar to the previous one. 
There is a single, highly linearly polarized component at all 
frequencies. The PA swing is rather flat but appears to steepen towards
the trailing edge of the profile.  Again, the pulse width decreases with
frequency as the trailing edge becomes sharper. There is a slow decline
in the fractional polarization with frequency.

\noindent
{\bf PSR~J0630$-$2834:} The profile of this pulsar consists of a single
broad component with a full width of $\sim$40\degr. The swing of PA
is steepest at the component centre at all frequencies. This indicates
that there is not much difference in emission height between the lowest
and highest frequencies.  The pulse width and linear
polarization decline slowly with increasing frequency, a trend which
continues up to 8.4~GHz.

\noindent
{\bf PSR~J0729$-$1836:} The profile consists of a narrow leading
component and a broader trailing component. The profile gets narrower
with increasing frequency and the polarization fraction declines
but otherwise shows little change.
The PA swing shows an orthogonal jump near the start of the profile and
has a linear slope with no obvious steepening. At 3.1~GHz the pulsar
is weak and the polarization has disappeared. 

\noindent
{\bf PSR~J0837+0610:} The pulse profile consists of two blended components
with an overall width of only $\sim$9\degr, a width which appears to
increase at higher frequencies. The leading component has the steeper 
spectral index. The linear polarization profile also largely remains
constant with frequency. The PA swing is highly frequency dependent.

\noindent
{\bf PSR~J0908$-$1739:} The pulse profile consists of a steep rising edge
and a shallower trailing edge with the hint of an extra trailing
component in the wings of the profile. There is little variation in
pulse structure or width as a function of frequency. Linear polarization
is seen against the leading edge at low frequencies and more towards
the centre of the profile at higher frequencies. The PA swing is unclear
due to the presence of orthogonal mode jumps. The emission is likely
to arise from the leading edge of the beam only (Lyne \& Manchester 1988).

\noindent
{\bf PSR~J0922+0638:} The profiles of this pulsar at various frequencies
have been discussed at length in \cite{wcl+99}.
In that paper, the authors noted
that the profile appeared to bifurcate at 435~MHz but that no lower
frequency profiles were available. We can confirm the presence of a low
frequency component here. In the figure we have aligned the profile on
the trailing edge which corresponds to a location of high linear
polarization and positive circular polarization. Already at 660~MHz a
hint of the bifurcated component can be seen, and this component starts
to dominate the profile at 243~MHz. In addition to this new component
there is also a pedestal-like leading component which is somewhat stronger
at lower frequencies. The polarization swing is shallow across the
polarized part of the profile. This is likely to be a partial cone where
only the trailing edge is seen. The pulse width declines strongly with
increasing frequency as the leading edge becomes sharper.

\noindent
{\bf PSR~J1507$-$4352:} The profile of this pulsar consists of a single
narrow component at all frequencies. There is some narrowing of the profile
towards higher frequencies. The linear polarization fraction shifts away
from the pulse centre towards the leading edge as the frequency increases
and the fractional polarization decreases.
The PA gradient is almost constant across the pulse.

\noindent
{\bf PSR~J1559$-$4438:} The profiles of this pulsar at 0.69~GHz and
above were discussed in detail in Johnston et al. (2007)\nocite{jkk+07}.
In that paper we remarked on the strong frequency evolution between
1.4 and 3.1~GHz. At 322~MHz the profile is similar to that at 690~MHz
although the circular polarization is almost absent. At 243~MHz some
scatter broadening is seen which flattens the PA swing and has reduced
the amount of linear polarization.

\noindent
{\bf PSR~J1645$-$0317:} This pulsar has a `classical' variation
of pulse profile with frequency. At low frequencies the profile consists
of a single component. At frequencies above 1~GHz, outrider components
are seen flanking the main component and these become very prominent
at 5~GHz (von Hoensbroech \& Xilouris 1997\nocite{hx97}).
The linear polarization remains low at all frequencies.
The PA swing is rather confusing. There is an orthogonal jump against
both the leading and trailing outriders and in the middle of the profile
the PA takes a strange dip from its otherwise slow increase.
At 243~MHz however, the PA swings steeply following the profile peak,
with a lag of about 1.5\degr. This is not seen
in the lower time resolution profile of \cite{gl98}.
The pulse width is very narrow but increases with increasing frequency
as the outriders appear.

\noindent
{\bf PSR~J1703$-$3241:} This is a beautiful example of a triple profile
pulsar. The central component has the steepest spectral index and is
much less prominent at high frequencies. The profile gradually narrows
with increasing frequency. The PA swing is steepest through
the centre of the profile. The linear polarization is moderately high
and the circular polarization changes hand in the centre of the profile.

\noindent
{\bf PSR~J1705$-$1906:} The profile of this pulsar consists of a main
pulse and an interpulse separated almost exactly by 180\degr\
(Biggs et al. 1988, Weltevrede et al. 2007\nocite{blh+88,wws07}).
The main pulse consists of two components, each highly polarized at low
frequencies but with the leading component less so at high frequencies.
The circular polarization is negative on the leading edge but changes
sign towards the trailing edge. The interpulse is barely visible at
low frequencies but reaches about half the amplitude of the main pulse
at higher frequencies. However, this amplitude again declines at
still higher frequencies. The interpulse is virtually 100\% polarized.
The width of the main pulse decreases with increasing frequency.

\noindent
{\bf PSR~J1709$-$1640:} The pulse profile does not vary much as a function
of frequency except for a slight narrowing. In the higher signal to
noise profiles, an orthogonal jump can be seen just after the centre
of the pulse profile.  At 3.1~GHz the PA swing
has lost its RVM nature and the polarized fraction is substantially lower
than at lower frequencies.

\noindent
{\bf PSR~J1731$-$4744:} At low frequencies the profile of this pulsar
consists of two components with the first having some linear and circular
polarization. As the frequency increases, the pulse width slowly decreases
and a more centrally located component emerges. At 3.1~GHz the central
component is quite prominent and it also disrupts the rather smooth PA
swing seen at lower frequencies. The prominence of the central component
at higher frequencies is contrary to that seen in other triple pulsars
and a further example of this is seen in the next object.

\noindent
{\bf PSR~J1733$-$2228:} The frequency evolution of this profile is
very interesting. At 243~MHz it consists of single dominant component
followed by a weak component. At 322~MHz a third, trailing component
appears in the profile. At 0.69 and 1.4~GHz, the central and trailing
components are more prominent relative to the leading component and the
profile looks like a classical triple profile with a swing of circular
polarization through the centre. At the highest frequency the profile
has become boxy. It is hard to tell from the PA swing where the magnetic
pole crossing might be located.

\noindent
{\bf PSR~J1735$-$0724:} The profile evolution of this pulsar mirrors that of
PSR~J1645$-$0317. At high frequencies the central component is flanked
by two outrider components which become much less prominent at lower
frequencies. As with PSR~J1645$-$0317, linear polarization is low and
the PA swing exhibits a curious deviation from a slow swing in the centre
of the profile, apparently related to the narrow component of linear
polarization seen clearly at 322 and 690~MHz.
The pulse width increases with frequency.

\noindent
{\bf PSR~J1740+1311:} The pulse profile of this pulsar has been 
presented with very high s/n at 408 and 1400~MHz by Weisberg et al. (1999)
and Weisberg et al. (2004). At the lower frequency, 5 clear components
can be seen (especially in linear polarization). The midpoint of the
profile also coincides with the steepest PA swing (Johnston et al. 2005).
The 3.1~GHz profile is similar to the 1.4~GHz profile except for a reduction
in the linear polarization. Kijak et al. appear to show the central
component re-appearing at 4.8~GHz.
At 243~MHz we still can observe the leading component whereas in the
227~MHz observations of \cite{hr86} it appears to be absent as is also
the case in the 102~MHz profile of \cite{ims89}. The profile width
appears to increase with frequency.
\begin{figure}
\includegraphics[height=0.7\textheight,angle=0]{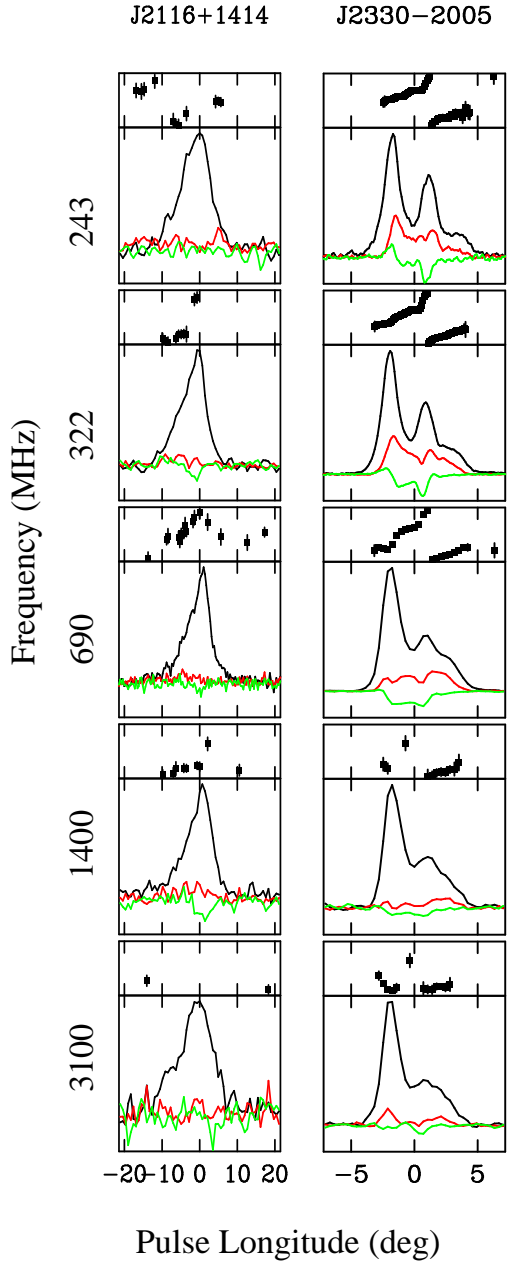}
\caption{Integrated profiles for 2 pulsars at 5 frequencies as marked.
See Figure~1 for details.}
\end{figure}

\noindent
{\bf PSR~J1745$-$3040:} The profile undergoes significant frequency evolution
which is similar to that in PSR~J1559$-$4438. At 3.1~GHz there is
a strong leading component, followed by a main component consisting of
two nearly equal components and a low ampltiude trailing component.
As the frequency decreases, so does the relative amplitude of the leading
component and the leading component of the main pulse. At 322~MHz
only the merest hint of the leading component is still present and
at 243~MHz a possible weak trailing component appears.
At high frequencies the PA swing is complex with an initial orthogonal
jump followed by a W shaped PA curve which has been distorted in the
pulse centre. The fraction of linear polarization also appears to
increase with increasing frequency.

\noindent
{\bf PSR~J1825$-$0935:} This peculiar pulsar has been the subject of much
debate in the literature. The profile consists of a double component
followed some 180\degr\ later by an interpulse and it appears as if
the interpulse and the leading component of the main pulse exchange
information (Gil et al. 1994\nocite{gjk+94}). Attempts have been made
to understand the geometry of this pulsar, most recently by \cite{dzg05}.
Although a 100~MHz observation of the main pulse has been published
(Izvekova et al. 1989\nocite{ims89}), no published profile of the interpulse
exists below 400~MHz. In both our 243 and 322~MHz profile the interpulse
is clearly present with the same separation from the main pulse as seen 
at higher frequencies.  At 243 and 322~MHz the leading component of the 
main pulse and the interpulse is very weak.  Although the 
leading component retains its polarization as the frequency increases, it 
decreases substantially in the trailing component.

\noindent
{\bf PSR~J1844+1454:} Strong profile evolution is seen in this pulsar.
At 3.1~GHz, the profile consists of a strong, narrow leading component
which is highly polarized followed by a weaker, broader trailing component 
with little polarization. At 1.4~GHz these components
are nearly equal amplitude. A steep swing of PA can be seen towards the
trailing edge of the profile. At 0.69~GHz the leading component is dominated
by the trailing component which completely takes over at yet lower
frequencies. As a result, the pulse width increases, reaching a maximum
at 1.4~GHz before decreasing again.

\noindent
{\bf PSR~1850+1335:} The profile consists of a single component at 
all frequencies. There is a small amount of pulse broadening at the
lower frequencies. The polarization is moderately high, except at 243~MHz
where little or no polarization is seen. There is a linear gradient
of PA across the pulse indicating a line of sight relatively far from
the magnetic axis. The pulse width decreases with frequency.

\noindent
{\bf PSR~1900$-$2600:} The profile of this pulsar shows significant
frequency evolution and there are clearly a large number of independent
components in the profile each with a different spectral index.
At the lowest frequency a strong central component dominates with weak
outriders flanking it. As the frequency increases the outriders become
stronger compared to the central component. At 1.4~GHz at least 5 different
components can be discerned and the profile has become boxy at 3.1~GHz.
The overall pulse width remains largely constant with frequency.
A swing in the sense of the circular polarization can be seen across the
centre of the profile at all frequencies. As expected for a complex
profile the PA swing is significantly disturbed from that expected in
any simple geometric model.

\noindent
{\bf PSR~1913$-$0440:} The profile of this pulsar evolves from a simple
single component at the lowest frequency to a triple component pulsar
at 3.1~GHz. The component at low frequencies can be most readily 
identified with the trailing component at high frequencies. The fraction
of linear polarization is low and the PA traverse is peculiar and
not consistent between frequencies. At 0.69 and 1.4~GHz it appears to
sharply rise towards the trailing edge of the profile. At 322 and 243~MHz,
in contrast, the PA swing appears to be flat across the trailing edge yet drops
sharply through the pulse centre.
\begin{figure*}
\includegraphics[height=0.7\textheight,angle=0]{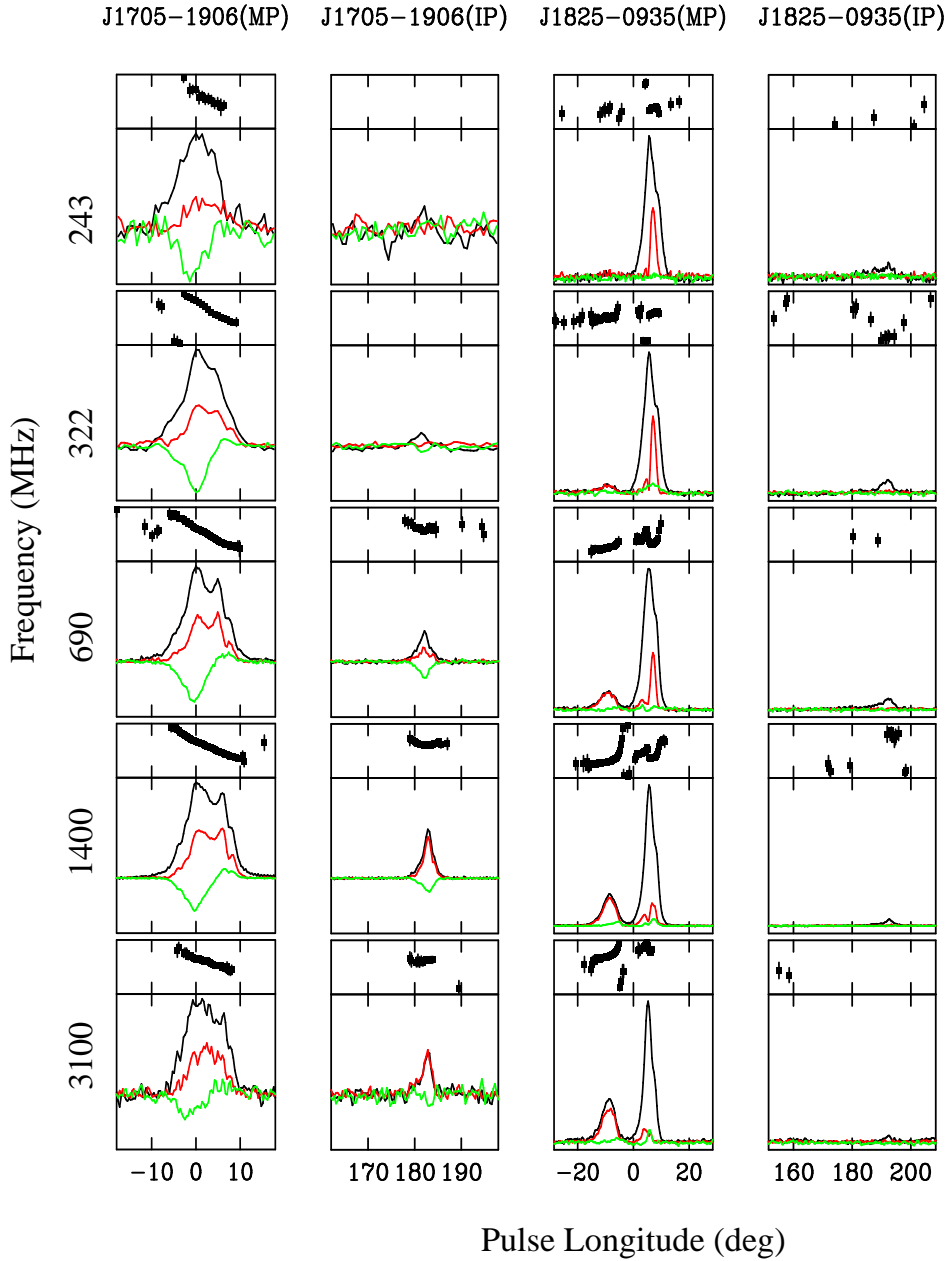}
\caption{Integrated profiles for the main pulse and interpulse of
PSR~J1705$-$1906 and PSR~J1825$-$0935 at 5 frequencies as marked.
The scale on the main and interpulse plots are the same.
See Figure~1 for details.}
\end{figure*}

\noindent
{\bf PSR~J1921+2153:} The profile of this, the first pulsar discovered,
is rather unusual. It consists of two blended components with likely
a central component also present. The spectral index of the outer two
components is similar and flatter than that of the central component which
has all but disappeared by 3.1~GHz. Weisberg et al. (1999)\nocite{wcl+99}
discuss the presence of a further leading component seen at 1.4~GHz which 
is also evident in our observations (see also Johnston et al. 2005).
However, this component is not present at any lower frequency. At 3.1~GHz,
however, in spite of the low s/n, it appears as if this leading component
has become stronger. The profile width is relatively constant with frequency.
The PA swing at all frequencies is complicted by
orthogonal mode jumps and component blending.

\noindent
{\bf PSR~1941$-$2602:} At 243~MHz the profile of this pulsar consists
of a dominant leading component and a weaker trailing component.
The spectral index of the trailing component is steeper than that
of the leading component so that, at 3.1~GHz the trailing component
is barely visible. The linear polarization appears to increase with
increasing frequency. The PA swing is flat throughout and the overall
width remains fairly constant. This may also be a partial cone, with
emission seen only from the leading edge of the beam.

\noindent
{\bf PSR~2048$-$1616:} The pulse profile of this strong pulsar has had
a variety of interpretations over the years (e.g. most recently by
Mitra \& Li 2004 and Johnston et al. 2007\nocite{ml04,jkk+07}). In brief,
the profile shows three major components, of which the central one
has the steepest spectral index. However, the central component is not
located mid-way between the two outer components but lies more on
the trailing side. It also lags the steepest swing of PA.
At 243~MHz there is a hint that a further central component is emerging
on the leading side.
The profile at 102~MHz in \cite{ims89} is peculiar as it only consists
of two components; a high signal-to-noise ratio profile at these low
frequencies would be very useful.  The pulse width decreases with
increasing frequency as does the fractional polarization.

\noindent
{\bf PSR~2116+1414:} The profile consists of a single component which
has a steeper trailing edge than a leading edge. The profile width remains
constant with frequency or perhaps increases slightly.
The fractional polarization is low at all frequencies.

\noindent
{\bf PSR~2330$-$2005:} The profile of this pulsar consists of three
components. At 243~MHz the leading component dominates followed by
the central and lagging component. The middle component has the steepest
spectral index and it is much less prominent at 3.1~GHz.
The overall pulse width is only marginally greater at 243~MHz than
at 3.1~GHz but appears to decrease monotonically with increasing frequency.
The linear polarization fraction is moderately high 
at low frequencies but has declined significantly by 3.1~GHz.
In \cite{jkk+07} we surmised that there was a steep swing of PA across
the central component rather than an orthogonal jump. This is borne out
by the 243~MHz PA traverse which shows a continuous rise in the PA.

\section{Discussion}
Although we did not attempt to choose a representative sample of pulsars
as part of our selection criteria, we can see from the pulse profiles that
many of the typical characteristics of pulsars are on display.
In this section, we compare these objects to the
various paradigms of pulsar phenomenology which have become integral
in any attempt to understand pulsar radio emission.

\subsection{Frequency Evolution of Profiles}
Clearly, the integrated profiles of pulsars vary as a function of frequency,
sometimes mildly, sometimes dramatically. In virtually all 
the pulse profiles presented here, however,
individual components are present at all frequencies to a greater or
lesser degree. In our sample, only PSRs J1645$-$0317, J1844+1454 and
J1913$-$0440 show features at high frequency not present at 243~MHz.

It has been evident for many years that some form of radius-to-frequency
mapping occurs within the pulsar magnetosphere and that lower frequencies
are emitted from higher in the magnetosphere than high frequencies.
However, in the range 0.1 to 10 GHz or so, the effect is not large
and the bulk of the radio pulsars emit from a rather narrow range of
altitudes from low down in the magnetosphere.
In contrast, the variation in height seen in different components of
a pulsar at a given frequency can be rather large with the outer components
having higher emission heights than more central components 
(Gupta \& Gangadhara 2003\nocite{gg03}).

Radius-to-frequency mapping should therefore be clearest in the simplest
profiles, without the complicating effects of individual component
spectral index. Indeed, in the simple profiles of our sample, a general 
narrowing of the pulse width of between 20 and 30 per cent is 
observed from 240 to 3100 MHz, as seen in other studies
(e.g. Thorsett 1991\nocite{tho91a}).
However almost one-third of our sample shows the
profile width increasing with frequency, opposite to that expected. An
explanation for this apparent anomaly can be sought in the
fact that components towards the outside of the profile increase in
relative amplitude with frequency, causing the apparent pulse
broadening (see Table~\ref{sources}). To investigate the true
wdith evolution with frequency it becomes necessary to trace the
profiles out to very low signal levels (Kijak \& Gil 1997\nocite{kg97}),
a task not possible with our dataset.

\subsection{Polarization Evolution}
In the standard picture of pulsar emission, the degree of polarization
generally decreases with increasing frequency. A physical explanation
of this phenomenon is not straightforward (von Hoensbroech,
Lesch \& Kunzl 1998\nocite{hlk98}) but may simply be related to the
fact that the high frequency radiation traverses more of the magnetosphere
than the low frequency emission.

Although the general decrease of linear polarization with frequency is
seen in many of our pulsars, we can identify further trends seen in
subgroups of sources. For instance, a number of highly polarized
profile components remain highly polarized at high frequencies
(e.g. for PSR J0922+0638).
In some pulsars, the change in the degree of
polarization is not monotonic with frequency (e.g. PSRs J0922+0638,
J1703$-$3241 and J1844+1454). Generally speaking, simple
profiles are more likely to be characterized by high linear polarization
than complex profiles and individual components within a given
pulsar can behave differently with frequency. 

We can interpret the depolarization of the emission in two possible ways,
the first of which is a geometric rather than a physical effect and the
second which involves orthogonally polarized modes. It is likely
that both effects are important because even in pulsars where no 
obvious orthogonal jumps are seen, depolarization is still observed
as the frequency increases.
In the first interpretation, the high degree of polarization arises from the 
fact that the emission is forced to travel along the field lines with a high
Lorentz factor, $\gamma$. The detected radiation has a duration proportional
to $1/\gamma$, typically shorter than the data sampling rate, so that
several emission bursts are (incoherently) summed in the detector.
This results in a decrease in the polarization fraction.
At high frequencies, which arise from lower in the pulsar magnetosphere,
the squeezing together of the field lines exacerbate this effect and
result in reduced degree of polarization. This is a geometrical effect, and
in principle, can be overcome by higher sampling rates.
The second idea, discussed in detail in \cite{kjm05}, involves the
interaction of the two orthogonally polarized modes of emission present
in the magnetospheric plasma (Manchester, Taylor \&
Huguenin 1975\nocite{mth75}). In our simple model, the two
orthogonally polarized modes had different spectral indices, and the
fractional polarization seen in a given component depended on the relative
strengths of the two modes. Some observational evidence in support of this
model comes from the fact that highly polarized pulsars (or components)
also appear to have flatter spectral indices (see also von Hoensbroech,
Kijak \& Krawczyk 1998\nocite{hkk98}).

In addition to the degree of linear polarization, the dependence of
the PA swing with frequency in our sample is very interesting.
In general, the PA traverse
is highly frequency dependent, whereas in the simple
geometrical explanation it should be independent of frequency
(Radhakrishnan \& Cooke 1969\nocite{rc69a}). The PA at a given longitude is 
therefore highly modulated by processes in the magnetosphere,
in particular the occurrence of orthogonal mode jumps and, most likely,
the presence of overlapping components at different emission heights.
A clear example of this is PSR~J1559$-$4438. At 322~MHz, the PA swing
is well behaved and the linear polarization is high. At 3.1~GHz, in contrast,
the linear polarization is lower, several new components are emerging and
the PA swing has become horribly complicated.
In contrast, pulsars with high levels of linear
polarization and generally simple profiles tend to show PA swings
which closer resemble the geometrical paradigm of the rotating vector model
(e.g. PSR~J0630$-$2834).

\subsection{Classification}
The most common pulse-profile classification schemes depend
on the number of components that constitute the profile, their
behaviour with frequency and their polarization characteristics.

A glance at the pulsars in our sample shows that they contain most of the
well-defined pulsar types. In particular, 8 pulsars (PSRs J0151$-$0635,
J0152$-$1637, J0304+1932, J0525+1115, J0729$-$1836, J0837+0610,
J1705$-$1906 (main pulse), J1921+2153) would be classifed as `double' in
the scheme of \cite{ran83}. Their profiles are all two-humped with each 
component being the same width, they show little profile evolution with 
frequency except for profile narrowing, they generally have low polarization 
and the PA swing is steepest at or near the profile centre.
In the main, they tend to be long period 
pulsars, with periods 1~s or above, they have low values of $\dot{E}$
and their profile width is less than 20\degr.
The simplest interpretation is that the observational profile 
arises from a cone of emission located at a low altitude ($<300$~km) and 
close to the last open field lines.

Other pulsars seem to fall into the `partial cone' category defined
by \cite{lm88}. These are PSRs J0134$-$2937, J0543+2329, J0614+2229,
J0908$-$1739, J0922+0638 and perhaps J1941$-$2602. These pulsars have
a sharp outer edge (either leading or trailing) and a smoother inner
edge which becomes steeper at high frequencies. Polarization is high
and the PA swing appears to steepen where the total intensity emission
runs out. The periods of these pulsars tend to be near 0.5~s, and they
also have spin-down energies intermediate between the highly energetic
young objects and the low $\dot{E}$ pulsars.
The interpretation of these pulsars are that their emission
arises from a cone which is only partially illuminated and so (by chance)
only the leading or trailing edge is seen. Interestingly, the measured pulse
widths of these pulsars are already larger than those of the doubles in
spite of the fact that only half the profile is seen. This implies their
emission arises from significantly higher in the magnetosphere than
those of the doubles.

Finally there are some pulsars which one might conveniently shoehorn
into some classification but which, for the most part, have bizarre enough
behaviour that many different interpretations could be made.
These are PSRs J1559$-$4038, J1733$-$2228, J1740+1311, J1745$-$3040,
J1844+1454 and J1913$-$0440. One feature which they all have in common is
that their PA swing contains many features and are highly distorted
compared to expectations from geometrical models. In the context of
the \cite{kj07} model, this arises because the multiple components in
these complex profiles all arise from different emission heights.

\subsection{Time Evolution of Pulse Profiles?}
We can draw several strands of the above sub-sections together.
Observationally it is known that the young highly energetic pulsars
(with $\dot{E}>10^{35}$~ergs$^{-1}$)
have simple, highly polarized profiles which are formed 
relatively high in the pulsar magnetosphere
(e.g. Johnston \& Weisberg 2006\nocite{jw06}).
There are no young, high $\dot{E}$ pulsars in this sample, mainly because
of the difficulty in observing them at low frequencies.
These pulsars lie at low galactic latitudes and have high dispersion
and scattering measures. The combination of high sky background
temperature and high scattering renders the majority of young
pulsars undetectable below $\sim$1~GHz.

In our sample here, we see that many pulsars have
complex profiles with many components, low to medium polarization 
and severe disruption of the PA swing especially in the centre of the
profiles. These have a mean $\dot{E}$ of 10$^{32.6}$~ergs$^{-1}$ and
also appear to originate from relatively high in 
the magnetosphere. Finally, at least a subset of the lowest $\dot{E}$ pulsars
have narrow double profiles, low polarization and a more geometric
PA swing. Their mean $\dot{E}$ is 10$^{32.1}$~ergs$^{-1}$ and emission
likely occurs low in the magnetosphere.

We therefore hypothesise a possible time evolution sequence for 
pulsar profiles, consistent with the ideas initially presented
in \cite{ran93} and subsequently modified in \cite{kj07}, and manifesting
itself as different profile types for different values of $\dot{E}$.
The high $\dot{E}$ pulsars have relatively simple profiles arising from
a single cone of emission high in the pulsar magnetosphere. As they spin down, 
the range of emission heights increases, producing complex profiles
(including partial cones) and non-geometrical PA swings. Finally, many
low $\dot{E}$ pulsars show profiles consistent with a single emission cone at
a rather low height.

A greater degree of linear polarization generally
arises from components located high in the magnetosphere.
For low $\dot{E}$ pulsars this generally implies low frequencies; hence the
polarization declines rapidly with increasing frequency. For high $\dot{E}$
pulsars, however, with higher emission heights, the profiles remain
polarized up to much higher frequencies.

\section{Conclusions}
The combination of the GMRT and Parkes capabilities has allowed us to
create a high time resolution, full Stokes parameterization of
34 pulsars at frequencies between 240 and 3100~MHz. We have presented
their profiles and provided a succint description of each pulsar's
characteristics.

The general `rules' of pulsar profiles are seen in these data; the
profiles narrow with frequency, outer components are more prominent
at higher frequencies and the polarization fraction declines with frequency.

We surmise that the emission in high $\dot{E}$ pulsars arises from a 
high height in the magnetosphere, that moderate $\dot{E}$ pulsars have 
highly complex profiles and PA swings due to a large range of 
possible emission heights and that at least a subset of low $\dot{E}$
pulsars have simpler emission beams from a relatively low height.

\section*{Acknowledgments}
We thank the staff of the GMRT and Parkes for their support of this project
and S. Kudale and A. Noutsos for help with the observations.
The Australia Telescope is funded by the Commonwealth of 
Australia for operation as a National Facility managed by the CSIRO.
The GMRT is run by the National Centre for Radio Astrophysics of the 
Tata Institute of Fundamental Research.

\bibliography{journals,modrefs,psrrefs,crossrefs} 
\bibliographystyle{mn2e}
\label{lastpage} 
\end{document}